\documentclass[aps,prl,notitlepage,twocolumn]{revtex4-1}
\usepackage{amsmath,amssymb}
\usepackage{dsfont}
\usepackage{graphicx}

\begin{document}

\newcommand{\bra}[1]    {\left\langle #1\right|}
\newcommand{\ket}[1]    {\left| #1 \right\rangle}
\newcommand{\tr}[1]     {{\rm Tr}\left[ #1 \right]}
\newcommand{\av}[1]     {\left\langle #1 \right\rangle}
\newcommand{\proj}[1]   {\ket{#1}\bra{#1}}
\newcommand{\Jx}   {\hat J_x}
\newcommand{\Jy}   {\hat J_y}
\newcommand{\Jz}   {\hat J_z}

\author{T. Wasak, P. Sza\'nkowski and J. Chwede\'nczuk}
\affiliation{Faculty of Physics, University of Warsaw, ul. Pasteura 5, PL--02--093 Warszawa, Poland}

\title{Interferometry with independently prepared Bose-Einstein condensates}

\begin{abstract}
  We show that it is possible to reach the sub shot-noise sensitivity of the phase estimation using two independently prepared Bose-Einstein condensates as an input of an interferometer.
  In this scenario, the quantum correlations between the particles, which are necessary to beat the shot-noise limit, arise from the indistinguishability of bosons.
  Allowing for atom number fluctuations independently in each condensate, we calculate the ultimate bound of the sensitivity. Our main conclusion is that even in presence of major atom number fluctuations,
  an interferometer operating on two independent condensates can give very high sensitivity. We also show that the estimation from the
  measurement of the number of atoms utilizes these quantum correlations. This observation, in context of recent measurement of the Fisher information
  in a many-body system [H. Strobel {\it et al}, Science {\bf 345}, 424 (2014)], opens the way towards the construction of a new type of an interferometer operating below the shot-noise limit.
\end{abstract}

\maketitle

Whenever the value of an unknown parameter $\theta$ is extracted from a series of experiments, the result is inevitably burdened by the uncertainty $\Delta\theta$.
If the system, which is the subject of measurement consists of $N$ unentangled particles, this uncertainty is bounded by the shot-noise limit (SNL),
where the precision scales as $\Delta\theta\propto\frac1{\sqrt N}$.
To overcome this limitation, it is necessary to use a properly entangled state \cite{giovannetti2004quantum,pezze2009entanglement}.
Preparation of such probes has been at the center of attention in the field of quantum interferometry in the recent years.
In the case of photonic interferometers, the entanglement between the photons is most commonly generated in the process of the parametric-down-conversion \cite{pdc1,pdc2}. The correlated pairs of photons
are a result of a non-linear interaction of the incedent laser pulse with a crystal. For atomic interferometers, the useful particle entanglement has been achieved by means of two-body
interactions present in ultra-cold systems. Usually, such correlations are associated with the spin-squeezing of a two-mode sample
\cite{kitagawa1993squeezed,wineland1994squeezed,esteve2008squeezing,appel2009mesoscopic,gross2010nonlinear,leroux2010orientation,riedel2010atom,chen2011conditional,berrada2013integrated}.
Alternatively, in a process which resembles the down-conversion,
the interactions drive the scattering of pairs of entangled atoms from a coherent source, such as a Bose-Einstein condensate (BEC)
\cite{collision_paris,lucke2011twin, twin_beam, cauchy_paris,twin_paris}.

In all these cases, the correlated many-body state is prepared in a dedicated procedure. Below we show, that the particle entanglement which arises solely from the indistinguishability of bosons, 
which is one of the fundamental laws of quantum mechanics, might be a resource for the sub shot-noise (SSN) interferometry. Our inquiry is supported by an effect \cite{knight_qo}, 
which has been observed in the experiment \cite{grangier}.
When a pair of bosons is put into two modes $a$ and $b$, in the second quantization the state reads
\begin{equation}\label{tf}
  \ket\psi=\ket{1}_a\ket{1}_b,
\end{equation}
which implies there is no entanglement between the modes. However, from the particle point of view, the wave-function of the first ($1^{\rm st}$) and the second ($2^{\rm nd}$) boson will
read $\ket\psi=\frac1{\sqrt 2}\left(\ket{1^{\rm st}}_a\otimes\ket{2^{\rm nd}}_b+\ket{1^{\rm st}}_b\otimes\ket{2^{\rm nd}}_a\right)$. This is a state, where the entangled parties are the particles
and the non-classical correlation is a result only of
the indistinguishability. Naturally, when the two modes are spatially separated, it is impossible to observe this entanglement. The situation
changes when the two particles simultaneously pass through a beam-splitter. Then, the bosonic statistics comes into play and at the output is a NOON state, which in the second
quantization reads
\begin{equation}\label{noon}
  \ket\psi=\frac1{\sqrt 2}\left(\ket{2}_a\ket{0}_b+\ket{0}_a\ket{2}_b\right).
\end{equation}
The beam-splitter extracts the entanglement between the modes from
the initial particle entanglement due to the indistinguishability \cite{identical_plenio}.
On the other hand, it acts on each particle independently,
so it does not change the amount of particle-entanglement.
This means that a twin-Fock (TF) state (\ref{tf}) is just as particle-entangled as the NOON state (\ref{noon}).
The presence of strong entanglement between the modes makes the NOON state an ideal candidate for SSN metrology if
$a$ and $b$ are identified with the two arms of an interferometer.

This example shows that (at least for two particles)
particle entanglement due to the indistinguishability is a viable resource for quantum interferometry. 
However, since the ultimate bound of the sensitivity is the Heisenberg limit (HL): $\Delta\theta\propto\frac1N$, which is
$\sqrt N$ better then the SNL, the gain from quantum correlations is of growing importance when $N$ gets big.
These observations motivate our inquiry: under which conditions, two
groups of bosons, which come from two independent sources and entangled solely due to indistinguishability, are useful for metrology? Or in other words: under which conditions
the entanglement due to indistinguishability is a sufficient resource for SSN sensitivity, eliminating the need for a dedicated entangling procedure.

The interferometric contrast is best when each cloud forms a BEC (or an equivalently coherent collection of bosons). 
In an ideal case, condensates are pure states of $\frac N2$ particles, which together form a TF state
\begin{equation}
  \ket\psi_{\rm tf}=\ket{\frac N2}_a\ket{\frac N2}_b.
\end{equation}
Such state, passing through the Mach-Zehnder interferometer (MZI), can potentially give the sensitivity $\Delta\theta\propto\frac{\sqrt2}N$, only $\sqrt2$ worse then the HL \cite{pezze_mzi}.
This is an ultimate precision
which can be reached with two independent groups of bosons. However, a realistic BEC is not a state with a fixed number of particles. Rather, it exhibits atom number fluctuations from shot to shot and
therefore must be described by a mixture
\begin{equation}\label{mix}
  \hat\varrho=\left(\sum_{N_a=0}^\infty P_a(N_a)\ket{N_a}\bra{N_a}\right)\otimes\left(\sum_{N_b=0}^\infty P_b(N_b)\ket{N_b}\bra{N_b}\right).
\end{equation}
Here $P_i(N_i)$ is a probability for having $N_i$ particles in the $i$-th condensate. 
These probabilities contain the complete information whether the state (\ref{mix}) is useful for quantum metrology.
It is the main goal of this manuscript to extract this information, and we begin by re-writing the Eq.~(\ref{mix}) in a following way
\begin{subequations}\label{dens}
  \begin{eqnarray}
    \hat\varrho&=&\sum_{N=0}^\infty\sum_{n=-\frac N2}^{\frac N2} P_a\left(\frac N2+n\right)P_b\left(\frac N2-n\right)\times\\
    &\times&\ket{\frac N2+n,\frac N2-n}\bra{\frac N2+n,\frac N2-n}.\label{kets}
  \end{eqnarray}
\end{subequations}
Here $n=\frac12(N_a-N_b)$ is the atom number difference, $N=N_a+N_b$ 
and $|\frac N2+n,\frac N2-n\rangle\equiv|\frac N2+n\rangle_a|\frac N2-n\rangle_b$.
This form of the density matrix is useful because the two-mode interferometric transformations are generated by the angular momentum operators
$\Jx=\frac12(\hat a^\dagger\hat b+\hat a\hat b^\dagger)$, $\Jy=\frac1{2i}(\hat a^\dagger\hat b-\hat a\hat b^\dagger)$ and $\Jz=\frac1{2}(\hat a^\dagger\hat a-\hat b^\dagger\hat b)$, which
do not couple states (\ref{kets}) with different $N$.

In order to asses the interferometric usefulness of the state (\ref{dens}) we refer to the quantum Fisher information (QFI, denoted here by $F_q$). It is a quantity,
which tells what is the ultimate sensitivity $\Delta\theta$, optimized over all possible estimation schemes. Thus the value of $F_q$ depends only on the state $\hat\varrho$ and the
$\theta$-dependent transformation representing the interferometer \cite{braunstein1994statistical}. According
to the Cram\'er-Rao lower bound (CRLB), the relation between the sensitivity and the QFI is \cite{holevo2011probabilistic}
\begin{equation}\label{crlb}
  \Delta\theta\geqslant\frac1{\sqrt m}\frac1{\sqrt{F_q}},
\end{equation}
where $m$ is the number of measurements.
In line with the definition of the SNL, the values of $F_q>N$ are attainable only for particle entangled states. When the total number of particles fluctuates (as is the case of
the state (\ref{dens})),
the SNL is redefined to $F_q=\bar N$, with $\bar N$ being the average number of particles in the system \cite{fluct_smerzi}.

A linear interferometer is represented by an evolution operator $\hat U=\exp\left[-i\theta\hat J^{(n)}\right]$, where $\hat J^{(n)}$ is a scalar product of a unit vector $\vec n$
and a vector $(\Jx,\Jy,\Jz)^T$ of the angular momentum operators.
For such transformation the QFI does not depend on $\theta$ \cite{braunstein1994statistical} and reads
\begin{equation}
  F_q=2\sum_{i,j}\frac{(\lambda_i-\lambda_j)^2}{\lambda_i+\lambda_{j}}\left|J^{(n)}_{ij}\right|^2.
\end{equation}
Here, $\lambda_i$ is the $i$-th eigen-value (with a corresponding eigen-vector $\ket i$) of $\hat\varrho$ and $J^{(n)}_{ij}=\bra i\hat J^{(n)}\ket j$.
For illustration, we take $\hat J^{(n)}=\Jx$, i.e.
$\hat U=\exp\left[-i\theta\Jx\right]$. Such a mode-mixing operation has been implemented in various
atomic interferometers \cite{lucke2011twin,smerzi_ob,gross2010nonlinear,riedel2010atom,gross2010nonlinear,berrada2013integrated}.
For this transformation, the $F_q$ calculated using (\ref{dens}) is
\begin{subequations}\label{fq}
  \begin{eqnarray}
    F_Q=\frac12\sum_{N=0}^\infty\sum_{n=-\frac N2}^{\frac N2}\left[\frac{\left(\lambda^{(N)}_n-\lambda^{(N)}_{n-1}\right)^2}{\lambda^{(N)}_n+\lambda^{(N)}_{n-1}}a^{(N)}_n+
      \right.\\
      \left.+\frac{\left(\lambda^{(N)}_n-\lambda^{(N)}_{n+1}\right)^2}{\lambda^{(N)}_n+\lambda^{(N)}_{n+1}}a^{(N)}_{n+1}\right],
  \end{eqnarray}
\end{subequations}
where $a^{(N)}_n=\left(\frac N2+n\right)\left(\frac N2-n+1\right)$ and $\lambda^{(N)}_n=P_a\left(\frac N2+n\right)P_b\left(\frac N2-n\right)$.
This expression allows for a quick estimate whether the state $\hat\varrho$ is useful for quantum metrology.
Although we used $\hat J^{(n)}=\Jx$, any transformation in the $x$-$y$ plane gives the same QFI for the state (\ref{dens}). For instance, when
$\hat J^{(n)}=\Jy$,  the following results would apply to the MZI.

\begin{figure*}[htb!]
  \includegraphics[clip, scale=0.5]{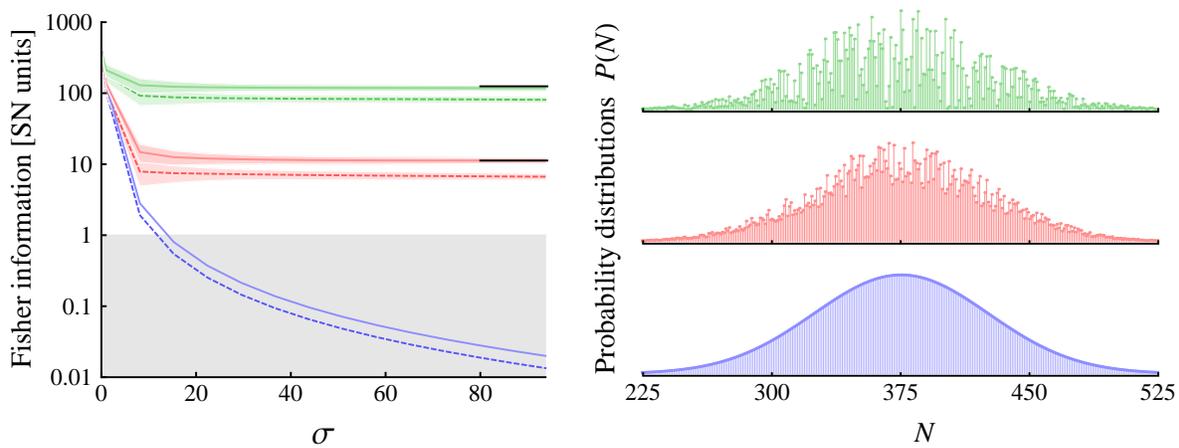}
  \caption{
    Left panel: The QFI (solid lines) and the Fisher information (dashed lines) for the population imbalance measurement as a function of the width $\sigma$ for $\bar N=750$ particles normalized to the SNL 
    and averaged over a statistical ensemble of disorders $\xi$, with statistical properties described in the text.
    In all cases, the shades show the related statistical widths. 
    From the bottom to the top: $\varepsilon=0,\ 0.3, 1$. For $\varepsilon=0.3$ and $\varepsilon=1$, short horizontal black lines at large $\sigma$ indicate the
    limiting values of the QFI calculated using Eq.~(\ref{limit}). Right panel: The exemplary probability distributions for $\sigma=50$.
  }\label{fig}
\end{figure*}

We now show that Eq.~(\ref{fq}) can take a particularly meaningful form under the following assumptions. When the probabilities $P_{a/b}$ change smoothly
over the $n\pm1$ increment, the difference between the two neighboring eigenvalues in Eq.~(\ref{fq}) can be approximated with a derivative. 
Consequently, sums change into integrals. When $P_{a/b}$ are peaked around the average number of
atoms in each BEC, then $a^{(N)}_n$ is a slowly varying function of $n$ and can be approximated with $\frac{N^2}4$. As a result, Eq.~(\ref{fq}) simplifies to
\begin{equation}\label{cont}
  F_q\simeq\frac18\int_0^\infty\!\! dNN^2\!\!\int_{-\frac N2}^{\frac N2}\!\!dn\,\frac{1}{\lambda^{(N)}_n}\left(\frac{\partial\lambda^{(N)}_n}{\partial n}\right)^2.
\end{equation}
These integrals can be estimated assuming that the probabilities are peaked around $\frac12\bar N$ each, with the spread equal to $\sigma$, giving a universal scaling
\begin{equation}\label{smooth}
  F_q\propto\frac{\bar N^2}{\sigma^2}.
\end{equation}
For instance, for Gaussian $P$'s \cite{stringari} we have $F_q=\frac{\bar N^2}{4\sigma^2}$.
This means that the state (\ref{dens}) is likely to provide SSN sensitivity when the atom number fluctuations in each BEC are sub-Poissonian \cite{fluct}.
To achieve this, one must be able to precisely control the process of condensation, i.e. manipulation of a single coherent cloud.
Note also that the quality of the state (\ref{dens}) is limited only by the life-time of each condensate, since there is no coherence between the two input arms of an interferometer.

It might happen that the atom number distribution is not a perfectly smooth function. To illustrate the impact of such imperfections,
we take $P_{a/b}$ as a common smooth envelope $P$ on top of which a stochastic process $\xi(N_i)$ models the disorder, namely
\begin{equation}\label{prob}
  P_{a/b}(N_i)\rightarrow P(N_i)\Big(1+\varepsilon\cdot\xi(N_i)\Big).
\end{equation}
Here $\varepsilon\in[0,1]$ is the amplitude of the disorder, while the noise is of the order of unity.
The condition that $P_{a/b}$ are equal could be satisfied if the two condensates were produced using some standardized experimental methods.
Nevertheless the main conclusion that follows is valid also when $P_{a/b}$ are disturbed by independent disorders and have different envelopes. 

For illustration, we numerically calculate the QFI using Eq.~(\ref{fq}) with the Gaussian envelope (peaked around $\frac12\bar N$ with a width $\sigma$) 
and the disturbance obtained
by drawing random numbers from the interval $[-1,1]$ independently for each $N_i$. The process $\xi$ generated in such a way is stationary, with a zero mean and is
characterized by the correlation function $\overline{\xi(N_i)\xi(N_j)}=\kappa(|N_i-N_j|)=\frac13\delta_{N_i-N_j,0}$.
The result is shown
in Fig.~\ref{fig} for $\varepsilon=0.3$ and $\varepsilon=1$ as a function of the width $\sigma$ with $\bar N=750$ particles.
Apparently, by introducing the disorder the ill-effect of particle number fluctuations has been reduced. Moreover, the larger amplitude of the disturbance gives higher QFI.


To provide a quantitative explanation of this unintuitive behavior, we refer to Eq.~(\ref{fq}). The value of the QFI depends on the difference between the neighboring eigenvalues of the density matrix.
Thus violent jumps between the adjacent $\lambda$'s increase the QFI.
This is also evident in the continuous limit (\ref{cont}) and is embodied by the derivative of $\lambda$ over $n$.
These observations explain why in the presence of the disorder 
the value of the QFI grows.
In line with these arguments, a disorder which is more smooth (i.e. when the range of the correlation function $\kappa$  extends over the neighbors) should have less beneficial influence on the QFI.

Moreover, in Fig.~\ref{fig} we observe that while in the absence of the disturbance the value of the QFI drops monotonically with $\sigma$, in presence of disorder 
it reaches some constant for major atom number fluctuations. This limiting value can be obtained analytically by averaging the QFI over different realizations of a stationary process $\xi$. 
Although, according to Eq.~(\ref{smooth}), the contribution from the smooth envelope tends to zero for large $\sigma$, the noisy part adds a $\sigma$-independent term 
\cite{supp} so that
\begin{equation}\label{limit}
  F_Q\propto\varepsilon^2\bar N^2(\kappa(0)-\kappa(1)).
\end{equation}
For the case used in Fig.~\ref{fig}, we obtain $F_Q\xrightarrow{\sigma\gg1}\frac{\varepsilon^2}{6}\bar N^2$ which is in excellent agreement with the numerical results. 
Also note that expression (\ref{limit}) confirms that a smoother disorder, when $\kappa(1)\simeq\kappa(0)$, gives a smaller contribution to the QFI.

We now demonstrate that a double-BEC system can give SSN sensitivity of the phase estimation from the population imbalance measurement. In the experiment we have in mind, the number of
atoms is measured in the output arms of the interferometer. For each total number of atoms $N$, the atom number difference $n$ is calculated and the
phase is deduced from the probability of obtaining $n$ given $\theta$, which reads
\begin{equation}\label{prob_imb}
  p_N(n|\theta)=\tr{\hat\varrho(\theta)\ket{\frac N2+n,\frac N2-n}\bra{\frac N2+n,\frac N2-n}},
\end{equation}
where $\hat\varrho(\theta)$ is the density matrix (\ref{dens}) propagated with the operator $e^{-i\theta\hat J_x}$. For this particular estimation scheme, the QFI in Eq.~(\ref{crlb})
must be replaced with the Fisher information, which reads
\begin{equation}\label{fish}
  F=\sum_{N=0}^\infty\sum_{n=-\frac N2}^{\frac N2}\frac1{p_N(n|\theta)}\left(\frac{\partial p_N(n|\theta)}{\partial\theta}\right)^2.
\end{equation}
Figure \ref{fig} compares the QFI with the Fisher information (\ref{fish}) optimized over the angle $\theta$ for a smooth Gaussian and distorted probabilities.
Although the estimation from the population imbalance is not optimal (it does not saturate the bound of the QFI) it still provides SSN sensitivity. 
The Fisher information (\ref{fish}) can be measured by detecting the distance between two neighboring probabilities
$p_N(n|\theta)$ and $p_N(n|\theta+\delta\theta)$, as recently experimentally demonstrated in \cite{smerzi_ob}.
Indeed, for small $\delta\theta$ the Hellinger distance \cite{hell} between two probabilities is
\begin{equation}
  1-\sum_{N=0}^\infty\sum_{n=-\frac N2}^{\frac N2}\sqrt{p_N(n|\theta)p_N(n|\theta+\delta\theta)}\simeq \frac{(\delta\theta)^2}8F.
\end{equation}
Implementation of this interferometric scheme would confirm that non-classical correlations due to indistinguishability are a resource for SSN interferometry.

\begin{figure}[htb!]
  \includegraphics[clip, scale=0.32]{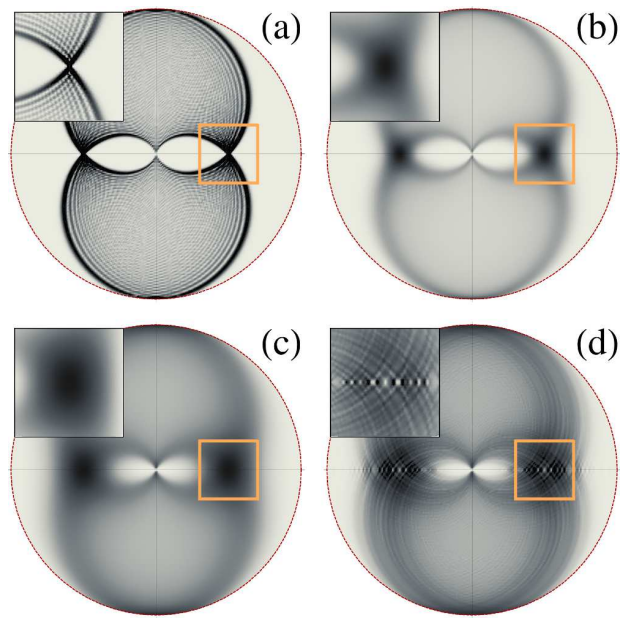}
  \caption{
    The probability distribution $p_N(n|\theta)$ from Eq.~(\ref{prob_imb}) calculated with Gaussian $P_{a/b}$ with mean $\bar N=100$ atoms and width equal to $\sigma=0.1$ (a), $\sigma=10$ (b)
    and $\sigma=20$ (c). 
    The panels show $p_{N=100}(n|\theta)$ as a function of 
    $n$ (radial variable) and $\theta\in[0,2\pi]$ (polar variable). The insets are enlargements of the regions marked with orange squares. When $\sigma$ grows (from (a) to (c)), fine structures
    of the probability vanish, giving a smaller value of the Fisher information (\ref{fish}). However, when the noise is added to the probability, according to Eq.~(\ref{prob}), 
    the fine strutures are restored, as shown in panel (d) for $\sigma=20$ and $\varepsilon=1$. 
  }\label{fig2}
\end{figure}

In this view, the Fisher information is related to the distinguishability of the probability distributions. 
A quantitative picture can be obtained by plotting $p_N(n|\theta)$ as a function of $n$ and $\theta$ with fixed $N$.
Figure \ref{fig2} shows such probabilities for different widths $\sigma$ of Gaussian $P_{a/b}$ and $\bar N=100$ particles and $N=100$. 
When $\sigma$ grows, fine structures of $p_{N=100}(n|\theta)$ diminish, and render this probability
at some $\theta$ less and less distinguishable from its neighbor at $\theta+\delta\theta$ \cite{pezze2009entanglement}. 
However, this trend is reversed by the addition of noise. The fine structures are restored, 
and in consequence the Fisher information grows, according to Eq.~(\ref{limit}). 


In conclusion, we have demonstrated that the particle entanglement between two independently prepared BECs might be a sufficient resource for SSN metrology. 
Such an input state is created in a process of condensation, which increases the coherence of each cloud, 
potentially giving high interferometric signal. Also, since initially there is no coherence between the two arms of the interferometer, 
the input state is not affected by the inter-mode decoherence and its quality is limited only by the life-time of each condensate. 

We have derived
the ultimate bound for the sensitivity of the parameter estimation. For smooth atom number distributions, it is possible to reach the SSN sensitivity when
fluctuations in each BEC are sub-Poissonian. Interestingly, for more erratic distributions, the sensitivity below the SSN can be reached even for vast atom number fluctuations.
Importantly, the precision of the phase estimation from the measurement of the number of particles exploits the particle entanglement in such system, opening the possibility for the experimental realization.
Our formulation provides all the necessary tools to evaluate the interferometric efficiency of any double-BEC configuration.

T.W. and P. Sz. acknowledge the Foundation for Polish Science International Ph.D. Projects Programme co-financed by the EU European Regional Development Fund.
This work was supported by the National Science Center grant no. DEC-2011/03/D/ST2/00200.

\bibliography{bibl}
\bibliographystyle{apsrev4-1}

\newpage
\clearpage

{\bf Appendix}

\bigskip{}

\setcounter{equation}{0} 
\makeatletter 
\renewcommand{\theequation}{A\@arabic\c@equation} 
\makeatother

In this Supplementary Material we derive the expression for the QFI with disordered probabilities. Following Eq.~(11) from the main text, we take 
\begin{equation}\label{prob}
  P_{a/b}(N_i)\rightarrow P(N_i)\Big(1+\varepsilon\cdot\xi(N_i)\Big).
\end{equation}
where $\varepsilon$ is the amplitude,  while the intensity of the noise is of order of unity. 
We take $\xi$ to be a stationary process so that its average, $\overline{\xi(N)}$ is constant, which without loss of generality can be set equal to zero. 
The correlation function depends only on the modulus of the argument difference, i.e. 
\begin{equation}
  \overline{\xi(N_i)\xi(N_j)}=\kappa(|N_i-N_j|)\,.
\end{equation}
Though the envelope $P(N_i)$ is a smooth function of $N_i$ with a width $\sigma$, the whole probability distribution might be erratic, because the disorder introduces rapid ``jumps''. 
The characteristic distance between these jumps, called the correlation length $N_\mathrm{corr}$, is described by the range of the correlation function $\kappa$,
\begin{equation}
  \kappa(\Delta N)\xrightarrow[\Delta N > N_\mathrm{corr}]{} 0\,.
\end{equation}
In the lowest order in $\varepsilon$, the QFI reads
\begin{subequations}\label{fq_dis}
  \begin{eqnarray}
    &&F_Q =F_Q^{(0)}+\varepsilon^2 \sum_{N=0}^\infty N^2\!\!\sum_{n=-\frac N2}^{\frac N2}\!\!\!\Delta\xi_n^{(N)}\times\\
    &&\times\left[\left(\lambda^{(N)}_n-\lambda^{(N)}_{n-1}\right)+\Delta\xi_n^{(N)}\lambda^{(N)}_n\right].
  \end{eqnarray}
\end{subequations}
Here, $\lambda^{(N)}_n=P\left(\frac N2+n\right)P\left(\frac N2-n\right)$, while $F_Q^{(0)}$ is the QFI in the absence of disorder and
\begin{subequations}
  \begin{eqnarray}
    \Delta\xi_n^{(N)}&=&\xi\left(\frac N2+n+1\right)-\xi\left(\frac N2+n\right)+\\
    &+&\xi\left(\frac N2-n+1\right)-\xi\left(\frac N2-n\right).
  \end{eqnarray}
\end{subequations}
The limiting value, to which the  QFI tends as $\sigma$ grows is given by the average of Eq.~(\ref{fq_dis}) over trajectories of the process $\xi$ and reads
\begin{subequations}
  \begin{eqnarray}
    &&\overline{F_Q}=F_Q^{(0)}+\varepsilon^2\sum_{N=0}^\infty\! N^2\!\!\!\sum_{n=-\frac N2}^{\frac N2}\!\![2\left(\kappa(0)-\kappa(1)\right)\\
    &&+\kappa(2n+2)-2\kappa(2n+1)+\kappa(2n)]\lambda^{(N)}_n.\label{der}
  \end{eqnarray}
\end{subequations}
It is now important to verify how the two terms proportional to $\varepsilon^2$ scale with $\sigma$. First, we take the continuous limit and  change the variables  
$\tilde n\equiv\frac n\sigma$ and $\tilde N\equiv\frac N\sigma$. Since the probabilities $P$ are peaked around $\frac12\bar N$, we obtain that a dominating contribution to the QFI at large $\sigma$ is
\begin{equation}\label{limit2}
  F_Q=F_Q^{(0)}+\mathrm{const}\times\varepsilon^2\bar N^2(\kappa(0)-\kappa(1)),
\end{equation}
where the constant comes from the double integration of the $\lambda$ function. Since the term in line (\ref{der}) in the continuous limit is the second derivative of the correlation function, it 
will scale inversely with $\sigma$ and is negligible for large atom number fluctuations. 

So far we have considered only the lowest order expansion of the QFI in $\varepsilon$. Note that the higher-order contributions, after taking the ensemble average, will be proportional to the integrals of
higher order correlation functions. All these functions have a characteristic range, which is of the order of $N_\mathrm{corr}$ \cite{kampen}, which is assumed to be small compared to $\bar N$. 
Therefore, these correlation functions will effectively act as Dirac delta's on top of the broad envelope set by $P$'s. As a result, the corresponding integrals will scale inversely with 
$\sigma$, giving a negligible contribution to the QFI for large atom-number fluctuations. 
This means that Eq.~(\ref{limit2}) is a universal expression for the QFI at large $\sigma$ for disordered potentials (\ref{prob}). 

Finally, we note that for a Gaussian envelope $P$ used in the main text, i.e.
\begin{equation}
  P(N) \propto \exp\left[-\frac{\left(N-\frac{\bar N}2\right)^2}{2\sigma^2}\right]
\end{equation}
we obtain
\begin{equation}
  \overline{F_Q} \approx\left(\frac{\bar N}{2\sigma}\right)^2 + \frac{\bar{N}^2\varepsilon^2}{2}\left[\kappa(0)-\kappa(1)\right],
\end{equation}
which is in excellent agreement with the numerical results presented in Fig.~1.

\end{document}